# Freestanding LiPON: from Fundamental Study to Uniformly Dense Li Metal Deposition Under Zero External Pressure

Diyi Cheng[1], Thomas Wynn[2], Bingyu Lu[2], Maxwell Marple[3], Bing Han[2], Ryosuke Shimizu[2], Bhagath Sreenarayanan[2], Jeffery Bickel[4], Peter Hosemann[4], Yangyuchen Yang[2], Han Nguyen[2], Weikang Li[2], Guomin Zhu[2], Minghao Zhang[2,\*], Ying Shirley Meng[1,2,5\*]

[1]Materials Science and Engineering Program, University of California San Diego, La Jolla, CA 92121, USA
[2]Department of NanoEngineering, University of California San Diego, La Jolla, CA 92121, USA
[3]Physical and Life Science Directorate, Lawrence Livermore National Laboratory, Livermore, CA 94550, USA
[4]Nuclear Engineering Department, University of California Berkeley, Berkeley, CA 94720, USA
[5]Pritzker School of Molecular Engineering, University of Chicago, Chicago, IL 60637, USA

*Corresponding Author E-mails: miz016@eng.ucsd.edu, shirleymeng@uchicago.edu

## Abstract

Lithium phosphorus oxynitride (LiPON) is a well-known amorphous thin film solid electrolyte that has been extensively studied in the last three decades. Despite the promises to pair with Li metal anode and various cathode materials, the presence of rigid substrate and LiPON's unique amorphous, air-sensitive nature set limitations to comprehensively understand its intrinsic properties for future development and applications. This work demonstrates a methodology to synthesize LiPON in a freestanding form that exhibits remarkable flexibility and a Young's modulus of ~33 GPa. Solid-state nuclear magnetic resonance (ss-NMR) and differential scanning calorimetry (DSC) results with unprecedented high signal-to-noise ratio could be obtained with such freestanding LiPON (FS-LiPON), revealing the Li/LiPON interface bonding environments quantitatively and a well-defined glass transition temperature for LiPON. Combining interfacial stress and a seeding layer, FS-LiPON demonstrates a uniform and fully dense Li metal deposition without the aid of external pressure. Such a FS-LiPON film offers new opportunities for fundamental study of LiPON material and associated interfaces, and provides perspectives for interface engineering in bulk solid-state battery.

# Introduction

Lithium phosphorus oxynitride (LiPON) is a thin film solid-state electrolyte (SSE) that is conventionally deposited on solid substrates. LiPON was first reported by Bates et al. in 1992 by substituting 5%-8% of O with N in $Li_3PO_4$ by radio-frequency (RF) sputtering in nitrogen.[1] LiPON rapidly drew research attention in the solid state battery field as it exhibits exemplary cyclability with a vast choices of electrode materials, i.e., $LiCoO_2$,[2] $LiMn_2O_4$,[2] $LiNi_{0.5}Mn_{1.5}O_4$,[3] $Li_4Ti_5O_{12}$,[4] and Li metal,[2–4] etc. Over the last three decades, numerous research efforts have been undertaken to shed light on the structure and properties of the material itself and associated interfaces, for the sake of gaining knowledge regarding the stable nature and performance of LiPON, and providing guidelines to engineering other SSEs for next generation lithium rechargeable batteries.

Interpretation of the intrinsic properties of LiPON, nevertheless, has been disputed due to the amorphous nature of LiPON, its sensitivity to ambient environment, and the presence of solid substrate. One of the long-lasting debates pertains to the N bonding structure and its impact on the lithium transport properties in LiPON. Early studies on the chemistry of LiPON primarily relied on X-ray photoelectron spectroscopy (XPS), where two different N environments were identified and regarded as double-bridging N and triple-bridging N.[5–7] Such peak assignment was in part due to the pervasive belief that the cross-linked configuration introduced by N incorporation is the key for the ionic conductivity enhancement of LiPON over $Li_3PO_4$.[7] Alternative methods suitable to probe local bonding environment such as neutron paired distribution function (PDF) and solid-state nuclear magnetic resonance (ss-NMR) were previously unable to validate this hypothesis due to the difficulty to obtain enough signal-to-noise (S/N) ratio, as the presence of substrates beneath LiPON thin film largely limits the active material amount that can be measured. Regardless, Lacivita et al. managed to obtain sufficient sample for neutron PDF measurements by scraping LiPON from the substrate and ruled out the existence of triple-bridging N, showing instead the prevalence of double-bridging N and apical-N bonding environments that was later verified with ss-NMR spectroscopy.[8,9]

Recent work has extended beyond bulk structure toward the interfacial stability between electrode materials and LiPON. Despite the knowledge on cathode-associated interfaces in early work[10–12] and recent insights gained on Li metal/LiPON interface via the advances of cryogenic electron microscopy (cryo-EM),[13,14] the electro-chemo-mechanical properties of SSE/electrode interfaces have yet to be explored. These are also regarded as critical metrics to determine the mechanical behavior at the interface during cycling, which can alter the stability and the cycle life of solid-state batteries.[15] However, literature studying the mechanical properties of LiPON has been limited. Herbert et al. measured the Young's modulus of LiPON by nanoindentation and obtained a value of 77 GPa.[16] A similar Young's modulus was documented by Xu et al. using picosecond ultrasonics measurement.[17] Such a modulus could account for its ability to suppress Li dendrite protrusion during cycling and contribute to its remarkable cyclability. Nevertheless, the long-term air exposure of LiPON during nanoindentation measurements and many approximations made during picosecond ultrasonics data analysis create some ambiguity. LiPON's film format also excludes mechanical testing beyond indentation (i.e. tensile, bending) due to presence of a substrate.

Such dilemma associated with the substrate and insufficient active material amount for measurements originates from the conventional synthesis methods of LiPON thin film. In fact, a variety of existing methods are available to synthesize LiPON, including RF sputtering,[1] pulsed laser deposition (PLD),[6] atomic layer deposition (ALD),[18] ammonolysis, inductively coupled plasma (ICP) and ball milling. Despite the diverse compositions of the LiPON variants yielded by different methods (**Supplementary Table S1**), most of these processes require the use of a solid substrate (i.e., silicon, glass, alumina, or sapphire, etc.) for LiPON to be deposited on, especially for vacuum deposition techniques. Those methods getting around the use of substrate such as ammonolysis, ICP or ball milling suffer either from the altered LiPON properties or the introduction of interfacial impedance between LiPON powders.[19–21] For the sake of enabling further fundamental characterizations, the removal of the substrate during LiPON synthesis will serve as a sound solution. Furthermore, substrate removal will also improve the energy density of thin film solid-state battery, as the substrate usually weighs several to hundreds of times more than the active materials.

Inspired by the semiconductor industry, where a photoresist is commonly used as a mask material that can withstand plasma-assisted deposition while being easily removed by organic solvents, herein we introduce a different methodology to synthesize a LiPON thin film that is in freestanding form without a rigid solid substrate. This freestanding LiPON (FS-LiPON) thin film exhibits transparency and a remarkable flexibility. The proposed methodology shows no modification to the LiPON structure, chemical bonding environments, and electronic properties, compared with the substrate-based analogues in literature. Leveraging this form factor, solid-state NMR of Li metal/FS-LiPON sample provides fresh quantitative insights on the interphase formation towards a better understanding of associated stability of Li/LiPON interface. Differential scanning calorimetry (DSC) on FS-LiPON sample gives a more accurate measurement on the glass transition temperature of 207 °C for LiPON. Mechanical testing demonstrates a Young's modulus of ~33 GPa and the remarkable flexibility of FS-LiPON films, opening up the chance of comprehensively studying LiPON's intrinsic mechanical properties using it freestanding form. All the above measurements benefit from the enhanced S/N ratio due to the removal of the substrate, illustrating the advantage and potential use of FS-LiPON for fundamental studies. With the aid of interfacial stress between at Cu/FS-LiPON and the presence of an Au seeding layer, we further demonstrate electrochemical deposition of uniform and fully dense Li metal under zero external pressure. The combination of interfacial stress and metal seeding layer for enabling uniform Li metal deposition provides new insights on interface engineering in bulk Li metal solid-state batteries.

## Results and Discussion

*A Flexible Freestanding LiPON Thin Film*

**Figure 1A** depicts the fabrication procedure of FS-LiPON. Before employing RF sputtering, a spin-coating method was used to coat a clean glass substrate with photoresist. Details about the spin coating recipe can be found in the experimental procedure section. LiPON thin film was then deposited onto the coated glass substrate by RF sputtering under $N_2$ plasma. After RF sputtering, LiPON sample was transferred into a container filled with Dimethyl carbonate (DMC) solvent in an argon-filled glove box. The substrate and LiPON film were fully immersed in DMC for overnight. Photoresist was then dissolved by DMC, after which LiPON film delaminated from the glass substrate and ready for pickup. Unlike the common way of producing LiPON thin film on a solid substrate, this method yields LiPON film in a freestanding form and exhibits transparency and remarkable flexibility as shown in **Figure 1B** and **Supplementary Video S1**. Such characteristics indicate LiPON being a soft material, in stark contrast with previous observations in literature.[16,17] Depending on the substrate size, deposition area and deposition time, the area, thickness and sample amount of FS-LiPON can be controlled following this procedure (**Supplementary Figure 1**).

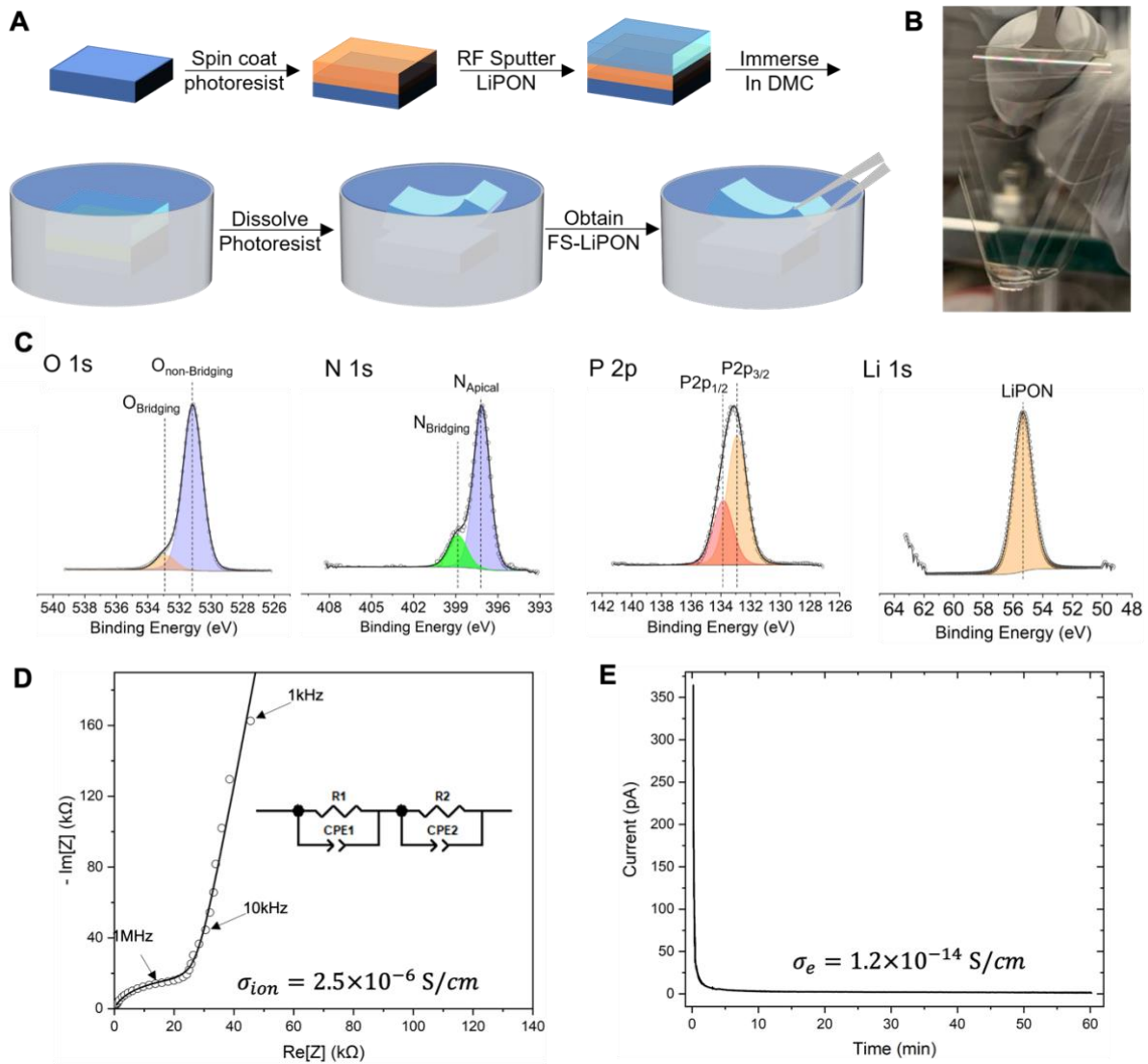

**Figure 1.** (A) Schematic of synthesis procedure for FS-LiPON. (B) Optical photo of a transparent and flexible FS-LiPON thin film. (C) XPS spectra of O 1s, N 1s, P 2p and Li 1s regions of FS-LiPON thin film. (D) EIS plot and (E) DC polarization plot of FS-LiPON.

A variety of characterizations are performed to ensure that the structure, chemical bonding environments and electrical properties of FS-LiPON are not affected during the above synthesis procedure. The cross-sectional scanning electron microscopy (SEM) and energy-dispersive X-ray spectroscopy (EDS) elemental mapping, shown in **Supplementary Figure S2,** demonstrate that FS-LiPON retains its fully dense nature in this 3.7-μm-thick film and that P, O, and N elements are uniformly distributed across the sample. **Supplementary Figure S3A** shows the X-ray diffraction (XRD) results of FS-LiPON, where no diffraction spot is present in the top diagram, and the integrated signal at the bottom only exhibits an amorphous feature around 23 degrees in 2θ, indicating the amorphous characteristic of FS-LiPON. **Figure 1C** displays the XPS result of FS-LiPON thin film, where O 1s, N 1s, P 2p and Li 1s regions manifest consistent features with substrate-based LiPON (Sub-LiPON) that are included in **Supplementary Figure S4, S5** and in literature,[7,22] illustrating that the chemical bonding environments are retained in the freestanding film. Moreover, the elemental mapping results by EDS on FS-LiPON surface in **Supplementary Figure S3B** confirm that N, P and O elements are uniformly distributed on the surface of the FS-LiPON film. Serving as an SSE, LiPON is an ionic conductor whilst an excellent electrical insulator. Electrochemical impedance spectroscopy (EIS) and direct-current (DC) polarization were subsequently employed to examine the electrical properties of FS-LiPON. The EIS spectrum in **Figure 1D** yields an ionic conductivity of $2.5\times10^{-6}$ S/cm for FS-LiPON, consistent with that of Sub-LiPON shown in **Supplementary Figure S6** and is better than the LiPON analogues produced by PLD and ALD methods.[6,18] DC polarization plot in **Figure 1E** gives an electronic conductivity of $1.2\times10^{-14}$ S/cm, on the order of Sub-LiPON as reported in literature.[7,23] Based on above results, FS-LiPON exhibits consistent properties with Sub-LiPON regardless of its freestanding form.

*New Opportunities for Fundamental Study of LiPON*

The Li/LiPON interface remains one of the most important interfaces in solid-state battery field, demonstrating extraordinary electrochemical stability.[3,13] As a model example to demonstrate the advantage of applying FS-LiPON for spectroscopic characterization, ss-NMR was performed on Li/FS-LiPON sample and the results are shown in **Figure 2A-C**. The Li/FS-LiPON sample was prepared by depositing Li metal on FS-LiPON film using thermal evaporation. **Figure 2A** shows the $^{31}$P magic angle spinning (MAS) NMR spectra of FS-LiPON and Li/FS-LiPON. A high signal-to-noise ratio of the NMR spectra was obtained, attributed to the increased sampling volume permitted by freestanding form of the samples. Based on the previous assignments on FS-LiPON,[9] four different structural units are identified in each spectrum, including orthophosphate tetrahedra $PO_4^{3-}$ ($Q^0_0$), $P_2O_7^{4-}$ dimers ($Q^1_0$), bridging-N $P_2O_6N^{5-}$ dimers ($Q^1_1$) and apical-N $PO_3N^{4-}$ units ($Q^0_1$). A clear difference regarding the content of these structural units is shown in **Figure 2B**. The Li/FS-LiPON sample shows an increase (13%) of $Q^0_0$ units relative to the FS-LiPON sample at the expense of PON units ($Q^0_1$ and $Q^1_1$). Such increase of $PO_4^{3-}$ content indicates that a large amount of $Li_3PO_4$ components were generated between Li metal and LiPON as a result of interface formation, consistent with our previous observation via cryo-EM.[13] The decrease of other structural units such as bridging-O configuration ($Q^1_0$), bridging-N configuration ($Q^1_1$) and apical-N configuration ($Q^0_1$) in turn facilitate the formation of interface components including $Li_3N$ and $Li_2O$ between Li metal and LiPON. $^7Li$ MAS NMR spectrum of Li/FS-LiPON in **Figure 2C** shows a clear shoulder around 7.5 ppm compared with FS-LiPON, indicating $Li_3N$ formation at the interface.[24] The slight peak shift shown in **Figure 2C** may be due to dynamical heterogeneities between the interfacial Li ions and Li ions deep in LiPON. Li metal was also clearly observed at 264 ppm in **Supplementary Figure S7**.

Previous observations by electron microscopy probed the spatial distribution of interface components between Li metal and LiPON,[13,14] while above ss-NMR results of Li/FS-LiPON sample provide quantitative insights on the content of interface components, revealing the amount of $Li_3N$ and $Li_3PO_4$ formation as the interface products. The peak shift in $^7Li$ MAS NMR spectra suggests an enhanced lithium kinetics in Li/FS-LiPON sample, likely contributed by the interface components such as $Li_3N$. The coupling of ss-NMR results with cryo-EM observation have depicted a more complete view of Li/LiPON interface both compositionally and spatially. As-formed interface components such as $Li_3N$, $Li_2O$ and $Li_3PO_4$ not only serves as a passivation layer to prevent LiPON from being continuously reduced by Li metal, but also improves the kinetics at the interface to potentially facilitate uniform lithium transport and nucleation.

The format of FS-LiPON also permits thermal property analysis. LiPON is known to be a glassy material for which the glass transition temperature is one of the most important metrics to determine its metastable states and application environments. Nevertheless, due to the limitation of active material for measurement, previously documented trials using DSC to examine the glass transition temperature of Sub-LiPON failed to capture clear transition behaviors.[25] To this end, DSC was conducted on FS-LiPON. Results in **Figure 2D** show an obvious glass transition with an onset temperature of 207 °C and inflection around 234 °C, consistent with LiPON glass transition temperature studied using spectroscopic ellipsometry.[26] Subsequent thermal response further captured the crystallization process of LiPON, along with the gas evolution observed during heating (inset images in **Figure 2D**). DSC results suggest that a proper temperature range to handle LiPON falls below 325 °C. Extra consideration needed to be taken when heat treatment is performed on LiPON-related samples.

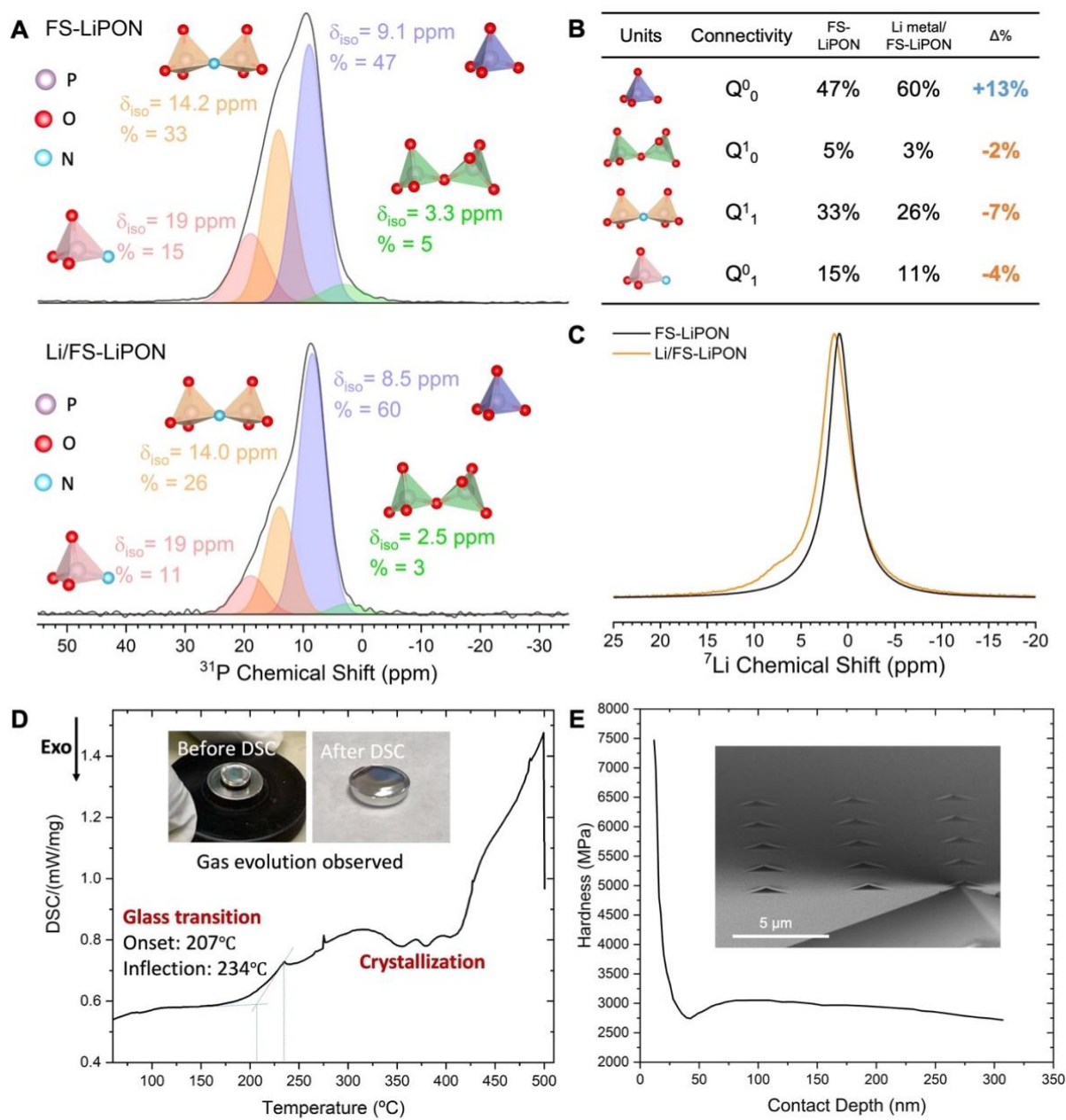

**Figure 2.** (A) $^{31}$P MAS NMR spectra of FS-LiPON and Li/FS-LiPON films. The spectrum of FS-LiPON was a reprint of our previous work[9], used here for comparison. (B) Structural unit component differences based on NMR deconvolution. $Q^0_0$ stands for the orthophosphate tetrahedra $PO_4^{3-}$ units, $Q^1_0$ stands for the bridging-O $P_2O_7^{4-}$ dimer units, $Q^1_1$ stands for the bridging-N $P_2O_6N^{5-}$ dimer units and $Q^0_1$ stands for the apical-N $PO_3N^{4-}$ units. (C) $^7$Li MAS NMR spectra of FS-LiPON and Li/FS-LiPON (D) DSC analysis of FS-LiPON film. The inset photos show the gas evolution due to DSC measurement. (E) Film hardness values measured via continuous stiffness measurement (CSM) indentation up to ~10% of the film thickness. Inset image shows the indents array on FS-LiPON during the nanoindentation experiment using a Berkovich indenter.

Outlooking the potential of FS-LiPON in the fundamental research, more chances are opened up to investigate the intrinsic properties of LiPON by mechanical tests. Note that it is essential to maintain an inert environment during the mechanical examination of LiPON, as **Supplementary Figure S8** and **Supplementary Video S2** show that FS-LiPON went through stiffening due to air exposure after 3 days. **Figure 2E** display the nanoindentation results collected within a vacuum chamber. The hardness values were plotted against contact depth of the indenter into FS-LiPON from continuous stiffness measurement (CSM), with statistics collected from over 100 indentation locations (inset image in **Figure 2E** shows the indents array at one sampling region). Nanoindentation gives an average hardness value around 2.7 GPa of FS-LiPON in the displacement range from 60 nm to 200 nm. The hardness values below 60 nm were primarily surface effect and have been excluded when determining the film hardness. Such a hardness of FS-LiPON is lower than the previously reported value of 3.9 GPa on sub-LiPON.[27] Based on the mathematical methods developed by Ma et al.[28] to determine Young's modulus from hardness, we obtained an average Young's modulus of FS-LiPON around 33 GPa, in contrast to the previously reported value of ~80 GPa for sub-LiPON.[27] It has been documented that vacuum deposition process commonly generates residual stress within thin film since the substrate may experience thermal expansion, contraction, or lattice mismatch, etc. during the deposition.[29] Specifically, sputtering process tends to generate compressive residual stress in the deposited thin film, which can affect the mechanical properties of thin films, such as increased hardness and Young's modulus.[30] Therefore, the removal of substrate for FS-LiPON has potentially resulted in stress release within LiPON film, which leads to diminished hardness and Young's modulus as seen. Such observation also suggests the importance to quantify residual stress in LiPON film before determining its mechanical properties.

As FS-LiPON film manifests a flexible feature, a series of flexibility testing were performed on FS-LiPON film to qualitatively understand the relationship between flexibility and film thickness. Results are shown in **Supplementary Video S3** and **Figure S9**, where films of varied thicknesses (1.7 μm – 3.7 μm) were taken for the bending test to examine the flexibility. A flathead tweezer was used to apply force on the FS-LiPON films whilst a video was taken to record the bending and breakage of the film. As the time-lapse series of images shown in **Figure S8**, all the FS-LiPON films exhibit remarkable flexibility upon bending. Right before film breakage, the 1.7-μm-thick film shows a high extent of bending compared with the 3.7-μm-thick film, indicating a higher flexibility.

Above results again illustrate the potential of using this freestanding film to obtain native properties of LiPON material itself. Regarding the mechanical properties, further tests such as tensile test, compression test, etc. can be performed on this FS-LiPON, which would otherwise be impossible to conduct on Sub-LiPON.

*Electrochemical Activity of FS-LiPON*

Apart from the intrinsic properties, FS-LiPON is also demonstrated to be applicable in electrochemical devices. A FS-LiPON Li-Cu cell was fabricated using the configuration shown in **Figure 3A**, where Cu and Li electrodes with the same designed area were aligned across FS-LiPON film. As-fabricated Li-Cu cell harnesses the flexible nature of FS-LiPON as shown in **Figure 3B**. The flexibility of the cell was further demonstrated in **Figure 3C**, where the cell was bent by the tweezer while still able to sustain Li metal plating and stripping capability afterwards. After cell fabrication, the cell was tested using the configuration in **Supplementary Figure S10**. **Figure 3D** shows the voltage curve of the Li-Cu cell during constant-current measurement. When a current of -50 nA is applied, the cell exhibits a voltage dip and reaches an overpotential of ~-1V, after which a stable plating process proceeds. When altering the current direction, a stripping curve feature is obtained. The cell demonstrated a stable plating and stripping over 13 cycles without short-circuiting, indicating the ability of FS-LiPON to shuttle lithium ions. The relatively high overpotential is likely caused by the resistance to deformation of the Cu current collector whist Li metal nucleates and grows. Apart from the nucleation barrier of Li metal, there is extra mechanical work needed to overcome the Cu deformation. It is noteworthy that owing to the unique configuration of FS-LiPON Li-Cu cell, no external pressure was applied to the cycled cell.

The plated Li metal morphology in FS-LiPON system was then examined by cryogenic focused ion beam/scanning electron microscopy (cryo-FIB/SEM). **Figure 3E** displays the cross-section morphology of pristine Li-Cu cell, where no extra layer is observed between Cu and FS-LiPON before plating and the evaporated Li metal on the other side of FS-LiPON appears fully dense. Note that cryogenic protection during FIB milling is crucial to preserve the pristine morphology and chemistry of Li metal, as reported elsewhere before[31] and demonstrated in **Supplementary Figure S11**. After a constant current plating, Li-Cu cell shows a dense Li layer with dark contrast between Cu and FS-LiPON in **Figure 3F**. Associated Li-Cu cell voltage curve is plotted in **Supplementary Figure S12**. EDS mapping on a plated

Li-Cu cell in **Supplementary Figure S13** illustrates the presence of Cu, P, O and Ga over corresponding regions. Due to the inability of regular EDS detectors to distinguish Li signal, the absence of EDS signal in the dense layers between Cu and FS-LiPON, and in the evaporated Li metal region suggests the existence of plated Li metal above FS-LiPON. Such features indicate a fully dense Li metal electrochemical deposition was realized by this FS-LiPON configuration when no external pressure was present.

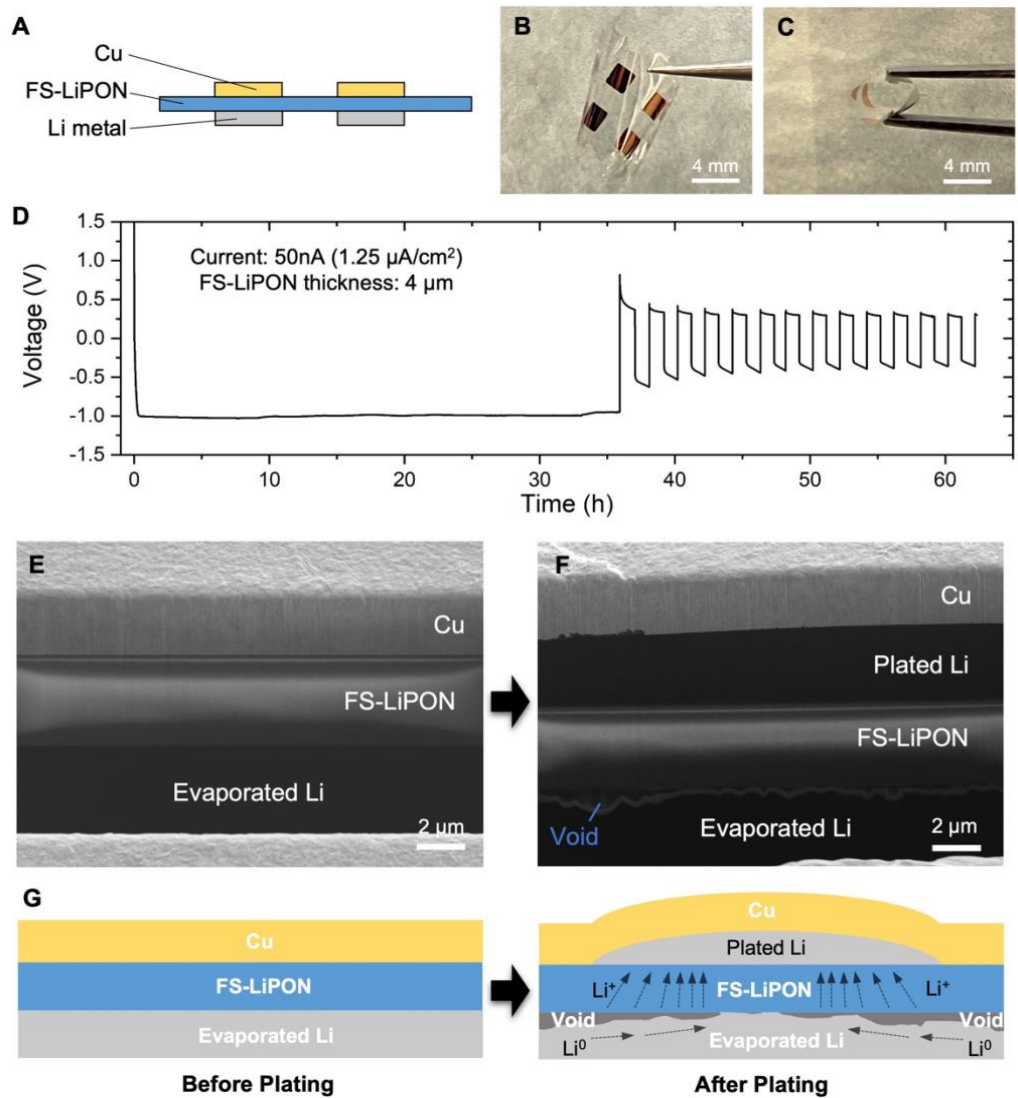

**Figure 3.** (A) Cross-section schematic of the FS-LiPON Li-Cu cell. Photos of FS-LiPON Li-Cu cell from top view (B) and upon bending (C). (D) Voltage curve of Li metal plating and stripping in a FS-LiPON Li-Cu cell. Cross-section cryo-FIB/SEM images of Li-Cu cell before Li metal plating (E) and after Li metal plating (F). The plated capacity in (F) is ~0.31 mAh/cm$^2$. (G) Schematic showing the proposed non-uniform void formation mechanism during Li metal plating.

Intriguingly, in **Supplementary Figure S13** a void region was observed between FS-LiPON and evaporated Li metal, as hinted by the aggregation of Ga signal that is commonly caused by redeposition during FIB milling and that is prevalently found at the bottom of void region after FIB milling.[13] A similar void feature is also observed in **Figure 3F**, where a gap presents between FS-LiPON and evaporated Li. Though the theoretical thickness of plated Li metal is calculated to be 1.5 μm based on the areal capacity, the observing region shows a plated Li metal thickness around 4 μm in **Figure 3F**. Top-view SEM image in **Supplementary Figure S14** displays various bumps distributed over the Cu surface after plating. **Figure 3G** delineates the plating process occurring in Li-Cu cell without pressure control. Before plating, each constituent in the cell is distinguishable by the well-defined interfaces. After plating, plated Li metal forces up the Cu layer around the initial nucleation site, while the non-uniform lithium-ion flux within FS-LiPON drives lithium atoms around the vicinity of the nucleation site to migrate and compensate the metallic lithium reservoir right under the nucleation site. Therefore, void regions are formed around the nucleation site after plating is completed. Above results suggest that Li metal plating is non-uniform across the FS-LiPON when cycling without external pressure. However, wherever it is plated between Cu and FS-LiPON, Li metal remains fully dense.

From a side view, the electrochemical deposition of Li metal in liquid electrolyte has long been problematic due to the uncontrollable mass transfer, non-uniform nucleation, and continuous growth of solid electrolyte interphase (SEI), which render the deposited Li metal torturous and wisker-like.[32,33] The electrochemically deposited Li metal in other SSE systems (i.e. $Li_7La_3Zr_2O_{12}$ and $Li_6PS_5Cl$, etc.), however, appears fully dense regardless of dendrite formation issues. Such morphological differences are likely due to the presence of external pressure on the order of several MPa in SSE systems.[34,35]

*Enabling Fully Dense, Uniform Li Deposition Without External Pressure*

Analogous to the cases of other solid electrolyte systems, fully dense Li metal plating has also been demonstrated in FS-LiPON system. Nevertheless, it is noteworthy that there is no external pressure applied on the LiPON system when the fully dense feature of plated Li is obtained, suggesting the possible presence of interfacial stress that could act as internal pressure to promote Li metal yielding and facilitate subsequent dense Li metal deposition. Previous work by Motoyama et al. proposed a model to simulate the interfacial stress between Li metal and Cu current collector after plating, where they used an imaginary Li metal sphere and estimated the radial stress on Li metal surface based on Hoop stress formula.[36]

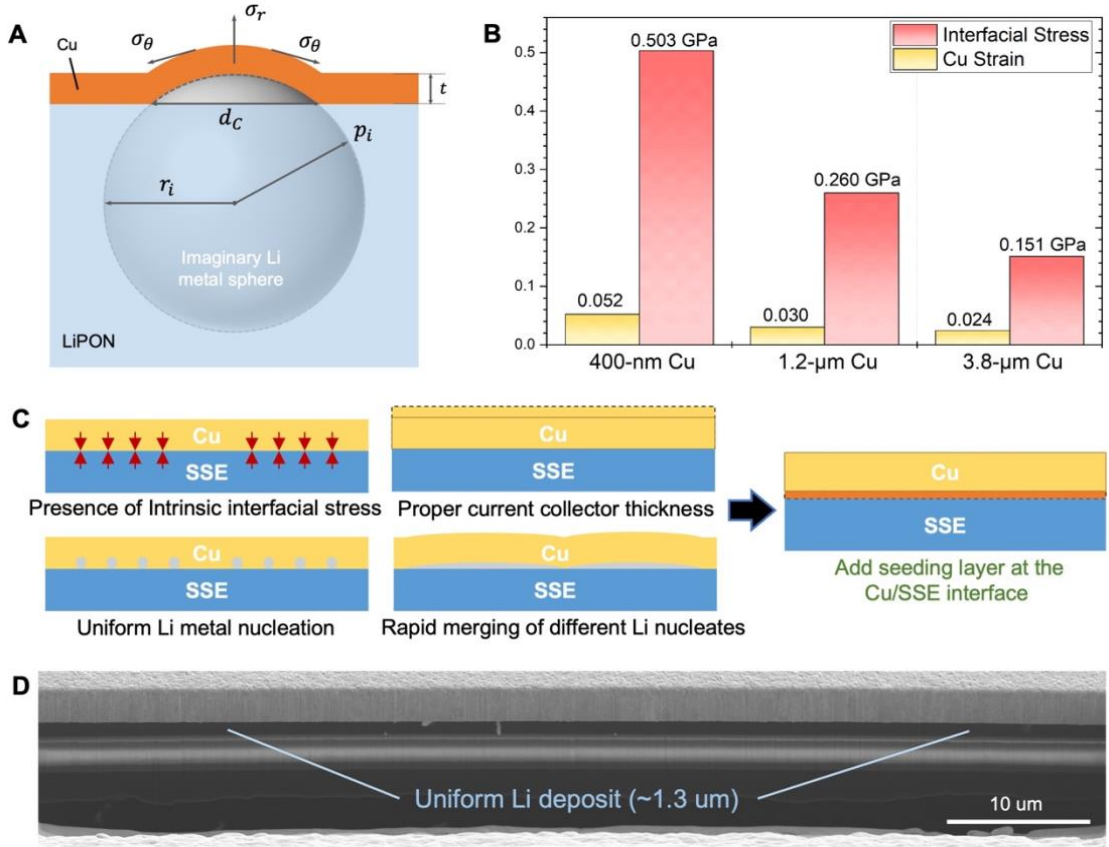

**Figure 4.** (A) Schematic of the interface model for interfacial stress simulation, where $P_i$ is the interfacial stress between Li and Cu, $\sigma_\theta$ is the stress on Cu in the circumferential directions, $\sigma_r$ is the stress on Cu in the radial direction, $t$ is the thickness of Cu, $d_c$ is the length of the chord marked in the Li sphere underneath the Cu dome region, $r_i$ is the radius of Li metal imaginary sphere. (B) Cu strain and simulated interfacial stress in Li-Cu cell with regard to different Cu thicknesses. (C) Proposed principles and solution to achieve uniform Li metal deposition in solid system. (D) Cryo-FIB/SEM image showing uniform Li deposition realized by adding Au seeding layer in FS-LiPON Li-Cu cell.

Employing the similar stress analysis model shown in **Figure 4A**, we obtained the formula as follows:

$$P_i = \frac{\varepsilon_{Cu} E_{Cu}}{(1 - \nu_{Cu})} \cdot \left\{ \frac{3(r_i + t)^3}{2[(r_i + t)^3 - r_i^3]} - \frac{\nu_{Cu}}{1 - \nu_{Cu}} \right\}^{-1}$$

where $P_i$ is the interfacial stress between Li and Cu, $\varepsilon_{Cu}$ is strain in the circumferential directions, $E_{Cu}$ is Young's modulus of Cu, $t$ is the thickness of Cu, $r_i$ is the radius of Li metal imaginary sphere, $\kappa$ is the curvature of Cu, $\nu_{Cu}$ is the Poisson's ratio of Cu. The input values of above parameters were extracted from **Supplementary Figure S15** and listed in

**Supplementary Figure S16**. **Figure 4B** shows the Cu strain and resulting interfacial stresses in the Li-Cu cells with different Cu thickness, ranging from 0.151 GPa to 0.503 GPa as Cu strain ramps from 0.024 to 0.052. Obtained stresses herein are hundreds of times higher than the external pressure applied on bulk SSE analogues. Such high interfacial stress present at Cu/Li interface confines Li metal morphology to achieve fully dense feature. Based on aforementioned stress formula, interfacial stress is inversely proportional to the Li deposit diameter and proportional to Cu strain, suggesting that Li metal deposit tends to have plenary growth so that overall stress can be released, resulting in more uniform coverage of Li metal on LiPON and less chance of dendrite formation. As such, we propose several criteria that need to be considered while building the ideal configuration for Li metal plating in solid state systems. As shown in **Figure 4C**, intrinsic interfacial stress is essential to generate pressure during Li metal plating without the aid of external pressure; proper current collector thickness is needed to confine Li metal morphology while maintaining its own structural integrity; uniform Li metal nucleation and rapid merging of different Li nuclei help reduce the plastic deformation of current collector to prolong cyclability. Consequently, one solution to achieve uniform Li metal deposition is adding seeding layer at Cu/SSE interface, as to facilitate uniform Li metal nucleation and subsequent uniform and dense Li metal growth. In this case, Au seeding layer was selected to demonstrate the hypothesis, since metal elements that alloy with Li metal tend to be lithophilic and can regulate nucleation behavior.[37] Prior to depositing Cu on FS-LiPON, a 3-nm-thick Au layer was first evaporated on FS-LiPON. Surface SEM image and EDS results in **Supplementary Figure S17** validates the Au film formed on FS-LiPON before Li-Cu cell fabrication. After electrochemical plating with zero external pressure (**Supplementary Figure S18A**), Cu surface remains relatively smooth as shown in **Supplementary Figure S18B**, suggesting a uniform Li metal deposition beneath. **Figure 4D** shows the cross-section image of the Li-Cu cell with Au seeding layer after plating. Measured thickness of Li metal deposit is ~1.3 μm, close to the thickness calculated from areal capacity (**Supplementary Figure S18A**). Li metal deposit appears not only full dense, but also uniform across the whole region. Small inclusions found in plated Li metal layer are likely the Li-Au alloy based on cross-section EDS in **Supplementary Figure S19**. Based on above results, with the aid of interfacial stress and seeding layer, uniform and fully dense Li metal deposition can be realized in solid-state system under zero external pressure.

Additionally, a further effort to demonstrate Li stripping in the Li-Cu FS-LiPON cell were summarized in **Supplementary Figure S20 and S21.** Although the Li-Cu FS-LiPON cell without external pressure ended up with non-uniform stripping that led to the formation of gaps between Cu and FS-LiPON and generates inactive lithium (**Supplementary Figure S20**), an external pressure of ~87.5 kPa helped largely improve the Coulombic efficiency to 82.7% (**Supplementary Figure S21**). It is important to stress that external pressure appears to be essential for the stripping process, while uniformly dense Li metal plating could be realized via interfacial stress and seeding layer solely. Another note is that the external pressure used in this set of experiments was nearly two orders of magnitude lower than the pressure commonly used in bulk solid-state Li metal batteries.[35,38] With such a low external pressure, a relatively good Coulombic efficiency for Li metal plating and stripping was achieved. An intriguing observation was that the gap/void caused by stripping was formed between Cu and Li metal instead of being present between Li metal and FS-LiPON (**Supplementary Figure S21C**), in contrast to the scenarios reported in bulk solid-state systems using argyrodite- or garnet-type SSEs.[39,40] Such difference is indeed related to the various current densities applied. Nevertheless, the fact that void was absent between Li metal and SSE in this case pertains to the unique characteristics of LiPON itself and Li/LiPON interface: LiPON is known for its amorphous nature, where grain boundaries are avoided. However, grain boundaries in other SSEs could be detrimental for creating current hotspot during Li plating/stripping as grain boundaries exhibit a higher electrical conductivity than bulk SSE does;[41] Meanwhile, the interface formed between Li and LiPON guarantees a facile lithium transport with excellent electrochemical stability that prevents current hotspot caused by continuous interfacial reactions. These features might have contributed to the absence of void formation between Li metal and LiPON and could shed light on the interface engineering in the bulk solid-state batteries.

## Conclusion and Outlook

This work presents a different methodology to produce a thin film SSE in a freestanding form that manifests transparency and remarkable flexibility. Basic characterizations validated the chemical compatibility of FS-LiPON against materials used during synthesis procedure. The absence of substrate for FS-LiPON largely leverages fundamental studies on LiPON material. Solid-state NMR illustrates fresh quantitative insights of Li/LiPON interface supplementing the previous findings by electron microscopy. DSC captures the glass transition behavior of LiPON around 207 °C with a high signal-to-noise ratio. Nanoindentation and flexibility test yield a Young's modulus of ~33 GPa of LiPON and show the flexible nature of LiPON film, respectively, calling for further mechanical tests to comprehensively explore LiPON's native mechanical properties. A further demonstration of an electrochemical cell employing FS-LiPON shows its ability to conduct lithium ions. Stress analysis at Li/Cu interface suggests the presence of a high compressive stress in the order of $10^{-1}$ GPa, which facilitates Li metal yielding and is the key for a dendrite-free, dense Li metal morphology. With the further aid of Au seeding layer, a fully dense and uniform Li metal deposition was realized under zero external pressure. The ideal conditions proposed for uniform Li metal deposition that combine interfacial stress and seeding layer provide new perspectives for interface engineering. The effort on Li metal stripping implies the essence of the amorphous nature of SSE and interfacial stability on preventing void formation within Li metal during the stripping process. With the freestanding form of LiPON thin films, opportunities have been opened up for a wider application of LiPON material. When coupled with casted cathodes, FS-LiPON can potentially be utilized as the SSE and enable Li metal anode with minimal external pressure.

# Experimental Procedures

## Photoresist Spin Coating

AZ1512 photoresist (from EMD Performance Materials Corp.) is coated on clean glass substrate by spin coating. The heater temperature for prebake and postbake was set as 100°C. The spinning recipe includes 500 RPM for 20s, 1000 RPM for 20s and 2000 RPM for 60s. The resulting photoresist layer thickness is about 1.7 μm.

## Thin Film Deposition

LiPON thin film was deposited on photoresist-coated glass substrate by RF sputtering using a crystalline $Li_3PO_4$ target (2″ in diameter, from Plasmaterials, Inc.) in UHP nitrogen atmosphere. The base pressure of the sputtering system was $3.0\times10^{-6}$ Torr. LiPON deposition used a power of 50W and nitrogen gas pressure of 15 mTorr. The as-deposited LiPON thin film was 3.7 μm in thickness with a growth rate of ~0.46 Å/s. The copper pads for EIS tests and current collector were deposited by thermal evaporation using copper pellets (from Kurt J. Lesker, 99.99% purity). Growth rate is 1 Å/s. Li metal anode for the Li-Cu cell was deposited by thermal evaporation with a base pressure of $2.5\times10^{-8}$ Torr and growth rate of 3-4 Å/s. The Au seeding layer was deposited by thermal evaporation using gold pellets (from Kurt J. Lesker, 99.99% purity). Growth rate is 1.5 Å/s.

## X-Ray Diffraction

The powder crystal X-ray diffraction was carried out on a Bruker micro focused rotating anode, with double bounced focusing optics resulting in Cu $K_{\alpha 1}$ and $K_{\alpha 2}$ radiation ($\lambda_{avg}$ =1.54178 Å ) focused on the sample. A sample of FS-LiPON was mounted onto a four circle Kappa geometry goniometer with APEX II CCD detector.

## Microscopic Morphology and Chemical Analysis

Scanning Electron Microscopy (SEM) was performed using an FEI Apreo SEM with an electron beam energy of 5 keV and an electron beam current of 0.1 nA. The energy dispersive spectroscopy X-Ray spectroscopy (EDS) was collected using an electron beam energy of 5 keV by the Pathfinder EDS software from Thermo Scientific.

## X-ray Photoelectron Spectroscopy

X-Ray photoelectron spectroscopy (XPS) was performed in an AXIS Supra XPS by Kratos Analytical. XPS spectra were collected using a monochromatized Al Kα radiation (hυ = 1486.7 eV) under a base pressure of $10^{-9}$ Torr. To avoid moisture and air exposure, a nitrogen filled glovebox was directly connected to XPS spectrometer. All XPS measurements were collected with a 300 × 700 μm$^2$ spot size. Survey scans were performed with a step size of 1.0 eV, followed by a high-resolution scan with 0.1 eV resolution, for lithium 1s, carbon 1s, oxygen 1s, nitrogen 1s, and phosphorous 2p regions. A 5 keV Ar plasma etching source was used for surface etching with a pre-etching for 5 s, etching for 60 s and post-etching for 10 s. All spectra were calibrated with adventitious carbon 1s (284.6 eV) and analyzed by CasaXPS software.

**Electrochemical Measurements**

A Biologic SP-200 potentiostat was used to measure the electrochemical impedance spectroscopy (EIS) and DC polarization of FS-LiPON, and electrochemical cycling of Li-Cu FS-LiPON cells. The frequency range for EIS was 3 MHz to 100 mHz with an amplitude of 10 mV and the obtained data fitted with a linear least square fitting method. The constant voltage used for DC polarization is 1V. The setup for electrochemical measurement is shown in the schematic in Supplementary Figure S6A and S10. To apply external pressure on Li-Cu FS-LiPON cell, a rigid stainless-steel plate (2×2×0.03 mm$^2$) was placed between the active region of the cell and the probe during measurements.

**Solid-state Nuclear Magnetic Resonance**

The NMR measurements performed on FS-LiPON and Li/FS-LiPON were collected using a 2.5 mm H/X/Y channel Bruker probe on a 600 MHz Bruker Biospin Avance III, operating at 242.94 and 233.25 MHz for $^{31}$P and $^{7}$Li. The samples were packed within a 2.5 mm pencil-type ZrO$_2$ rotor and spun at 25 kHz. The $^{31}$P spectra were collected as a rotor synchronized Hahn echo experiment with a 90° pulse of 2.54 μs (B$_1$ field strength ~98 kHz). The Hahn echo experiments were processed from the top of the echo to remove the effects of ring down from the FID. A single pulse experiment with a pulse length of 2.875 μs (B$_1$ field strength ~87 kHz) was used to acquire the $^{7}$Li spectra. The recycle delays used for the 1D experiments was 60 s for $^{31}$P and 2 s for $^{7}$Li.

**Differential Scanning Calorimetry**

The Differential Scanning Calorimetry (DSC) measurement was conducted with DSC 214 Polyma (Netzsch). The temperature range was from 50°C to 500°C with a heating rate of 10°C/min. The DSC measurement was conducted under $N_2$ environment. All samples were sealed in aluminum pans in an Argon-filled glovebox to reduce contamination.

**Nanoindentation**

Nanoindentation was performed inside of a Thermo Fisher Scientific Scios 2 DualBeam FIB/SEM using a FemtoTools FT-NMT04 nanoindenter equipped with a Berkovich tip. Measurements of hardness and reduced modulus employed the continuous stiffness measurement (CSM) technique using a displacement-controlled test. Mechanical property values were averaged between displacements of ~60 nm and ~200 nm. Several five-by-five indent arrays were performed at various locations on FS-LiPON films that were bonded to SEM stubs using epoxy. Tests were performed using a 4 s load-ramp time and a 0.2 s unload-ramp time. When transferring samples from an inert environment to the Scios 2 SEM, samples were exposed to <120s of atmosphere prior to the vacuum conditions inside the SEM.

**Cryogenic Focused Ion Beam/Scanning Electron Microscopy**

A FEI Scios DualBeam FIB/SEM equipped with cryo-stage was used to observe the surface and cross-section morphology of plated Li metal in FS-LiPON Li-Cu cell. The operating voltage of electron beam was 5 kV. Emission current of electron beam was set to 25 pA to minimize potential damage of electron beam. A gallium ion beam source was used to mill and thin the sample. The operating voltage of ion beam source was 30 kV. Emission currents of ion beam were chosen for different purposes, i.e., 10 pA for imaging by ion beam, 0.1 nA for cross-section cleaning and 3 nA for pattern milling. To preserve the Li metal pristine morphology, a cryo-stage was used during pattern milling and cross-section cleaning processes, where the temperature of cryo-stage was maintained at around -185°C due to heat exchanging with cooled nitrogen gas.

## Acknowledgements

The authors gratefully acknowledge funding support from the U.S. Department of Energy, Office of Basic Energy Sciences, under Award Number DE-SC0002357. FIB/SEM This work was performed in part at the San Diego Nanotechnology Infrastructure (SDNI) of UCSD, a member of the National Nanotechnology Coordinated Infrastructure, which is supported by the National Science Foundation (Grant ECCS-2025752). NMR was performed under the auspices of the US Department of Energy by LLNL under contract number DE-AC52-07NA27344. XPS and DSC were performed at the UC Irvine Materials Research Institute (IMRI) using instrumentation funded in part by the National Science Foundation Major Research Instrumentation Program under grant no. CHE-1338173 and DMR-2011967.

## Author Contributions

D.C., M.Z. and Y.S.M conceived the ideas. D.C., T.W., B.L., R.S. and B.S. prepared the thin film sample. The FS-LiPON Li-Cu cell was designed by D.C. B.H., M.Z. and G.Z., and fabricated by D.C. M.M. performed and analyzed ss-NMR measurements. D.C. conducted cryo-FIB/SEM and electrical measurements. D.C. and H.N. collected XRD data. D.C., J.B. and P.H. collected and analyzed the nanoindentation data. D.C., Y.Y. and W.L. collected XPS data. D.C., M.Z., Y.S.M., T.W., M.M., G.Z. and B.H. co-wrote the manuscript. All authors discussed the results and commented on the manuscript. All authors have approved the final manuscript.

## Declaration of Interests

The authors declare no competing interests.